\documentclass{llncs}
\usepackage{listings}
\usepackage{tabularx}
\usepackage{tikz}
\usetikzlibrary{patterns}
\usepackage{wrapfig}
\usepackage{pgfplots}
\usepackage{datatool}
\usepackage{float}
\usepackage{etoolbox}
\DTLloaddb[
headers={x,y,xx,yy,gg},
keys={x,y,xx,yy,gg}
]{data}{dataset_dist.dat}
\usepackage[linesnumbered, ruled, vlined]{algorithm2e}
\usepackage{array}
\newcolumntype{P}[1]{>{\centering\arraybackslash}p{#1}}

\usepackage[normalem]{ulem}

\title{Active Learning for Efficient\\Testing of Student Programs\textsuperscript{1}}
\author{Ishan Rastogi \and Aditya Kanade \and Shirish Shevade }

\institute{Department of Computer Science and Automation,\\ 
	Indian Institute of Science, Bangalore, India\\
	\email{\{ishanr, kanade, shirish\}@iisc.ac.in } }

\makeatletter
\def\blfootnote{\xdef\@thefnmark{}\@footnotetext}
\makeatother

\begin{document}
\begin{filecontents*}{dataset_dist.dat}
x,y,xx,yy,gg
0.0,5.9,0.0710526315789,6.1,1
0.0710526315789,5.9,0.09,6.1,0
0.1,5.9,0.150526315789,6.1,1
0.150526315789,5.9,0.19,6.1,0
0.2,5.9,0.250526315789,6.1,1
0.250526315789,5.9,0.29,6.1,0
0.3,5.9,0.364736842105,6.1,1
0.364736842105,5.9,0.39,6.1,0
0.4,5.9,0.448947368421,6.1,1
0.448947368421,5.9,0.49,6.1,0
0.5,5.9,0.558421052632,6.1,1
0.558421052632,5.9,0.59,6.1,0
0.6,5.9,0.663157894737,6.1,1
0.663157894737,5.9,0.69,6.1,0
0.7,5.9,0.745789473684,6.1,1
0.745789473684,5.9,0.79,6.1,0
0.8,5.9,0.869473684211,6.1,1
0.869473684211,5.9,0.89,6.1,0
0.9,5.9,0.968571428571,6.1,1
0.968571428571,5.9,0.99,6.1,0
0.0,4.9,0.063,5.1,1
0.063,4.9,0.09,5.1,0
0.1,4.9,0.148,5.1,1
0.148,4.9,0.19,5.1,0
0.2,4.9,0.252,5.1,1
0.252,4.9,0.29,5.1,0
0.3,4.9,0.357,5.1,1
0.357,4.9,0.39,5.1,0
0.4,4.9,0.452,5.1,1
0.452,4.9,0.49,5.1,0
0.5,4.9,0.55,5.1,1
0.55,4.9,0.59,5.1,0
0.6,4.9,0.65,5.1,1
0.65,4.9,0.69,5.1,0
0.7,4.9,0.75,5.1,1
0.75,4.9,0.79,5.1,0
0.8,4.9,0.85,5.1,1
0.85,4.9,0.89,5.1,0
0.9,4.9,0.95,5.1,1
0.95,4.9,0.99,5.1,0
0.0,3.9,0.0660638297872,4.1,1
0.0660638297872,3.9,0.09,4.1,0
0.1,3.9,0.16414893617,4.1,1
0.16414893617,3.9,0.19,4.1,0
0.2,3.9,0.267021276596,4.1,1
0.267021276596,3.9,0.29,4.1,0
0.3,3.9,0.369893617021,4.1,1
0.369893617021,3.9,0.39,4.1,0
0.4,3.9,0.46414893617,4.1,1
0.46414893617,3.9,0.49,4.1,0
0.5,3.9,0.572765957447,4.1,1
0.572765957447,3.9,0.59,4.1,0
0.6,3.9,0.662234042553,4.1,1
0.662234042553,3.9,0.69,4.1,0
0.7,3.9,0.768936170213,4.1,1
0.768936170213,3.9,0.79,4.1,0
0.8,3.9,0.869893617021,4.1,1
0.869893617021,3.9,0.89,4.1,0
0.9,3.9,0.976363636364,4.1,1
0.976363636364,3.9,0.99,4.1,0
0.0,2.9,0.0511111111111,3.1,1
0.0511111111111,2.9,0.09,3.1,0
0.1,2.9,0.147222222222,3.1,1
0.147222222222,2.9,0.19,3.1,0
0.2,2.9,0.241111111111,3.1,1
0.241111111111,2.9,0.29,3.1,0
0.3,2.9,0.353888888889,3.1,1
0.353888888889,2.9,0.39,3.1,0
0.4,2.9,0.436111111111,3.1,1
0.436111111111,2.9,0.49,3.1,0
0.5,2.9,0.544444444444,3.1,1
0.544444444444,2.9,0.59,3.1,0
0.6,2.9,0.64,3.1,1
0.64,2.9,0.69,3.1,0
0.7,2.9,0.735555555556,3.1,1
0.735555555556,2.9,0.79,3.1,0
0.8,2.9,0.843888888889,3.1,1
0.843888888889,2.9,0.89,3.1,0
0.9,2.9,0.941204819277,3.1,1
0.941204819277,2.9,0.99,3.1,0
0.0,1.9,0.0531081081081,2.1,1
0.0531081081081,1.9,0.09,2.1,0
0.1,1.9,0.152702702703,2.1,1
0.152702702703,1.9,0.19,2.1,0
0.2,1.9,0.252297297297,2.1,1
0.252297297297,1.9,0.29,2.1,0
0.3,1.9,0.355945945946,2.1,1
0.355945945946,1.9,0.39,2.1,0
0.4,1.9,0.451486486486,2.1,1
0.451486486486,1.9,0.49,2.1,0
0.5,1.9,0.550675675676,2.1,1
0.550675675676,1.9,0.59,2.1,0
0.6,1.9,0.652297297297,2.1,1
0.652297297297,1.9,0.69,2.1,0
0.7,1.9,0.751891891892,2.1,1
0.751891891892,1.9,0.79,2.1,0
0.8,1.9,0.853108108108,2.1,1
0.853108108108,1.9,0.89,2.1,0
0.9,1.9,0.955291479821,2.1,1
0.955291479821,1.9,0.99,2.1,0
0.0,0.9,0.045,1.1,1
0.045,0.9,0.09,1.1,0
0.1,0.9,0.143299595142,1.1,1
0.143299595142,0.9,0.19,1.1,0
0.2,0.9,0.241842105263,1.1,1
0.241842105263,0.9,0.29,1.1,0
0.3,0.9,0.337044534413,1.1,1
0.337044534413,0.9,0.39,1.1,0
0.4,0.9,0.439473684211,1.1,1
0.439473684211,0.9,0.49,1.1,0
0.5,0.9,0.539109311741,1.1,1
0.539109311741,0.9,0.59,1.1,0
0.6,0.9,0.639534412955,1.1,1
0.639534412955,0.9,0.69,1.1,0
0.7,0.9,0.738259109312,1.1,1
0.738259109312,0.9,0.79,1.1,0
0.8,0.9,0.837044534413,1.1,1
0.837044534413,0.9,0.89,1.1,0
0.9,0.9,0.936072874494,1.1,1
0.936072874494,0.9,0.99,1.1,0
\end{filecontents*}

\maketitle

\begin{abstract}
	
	In this work, we propose an automated method to identify semantic bugs in student programs, called ATAS, which builds upon the recent advances in both symbolic execution and active learning. Symbolic execution is a program analysis technique which can generate test cases through symbolic constraint solving. Our method makes use of a reference implementation of the task as its sole input. We compare our method with a symbolic execution-based baseline on $6$ programming tasks retrieved from CodeForces comprising a total of $23$K student submissions. We show an average improvement of over $2.5$x over the baseline in terms of runtime (thus making it more suitable for online evaluation), without a significant degradation in evaluation accuracy.
\end{abstract}

\section{Introduction}
\blfootnote{\textsuperscript{1} This is an extended version of paper due to appear in AIED 2018}
Recent times have seen a rise in the popularity of massive open online courses (MOOCs), which are attended by hundreds of students. This necessitates the development of automatic feedback generation techniques since human-based feedback may be prohibitively expensive, if not impossible. Owing to this, a number of automated feedback generation techniques for computer programming have been proposed in the recent literature. These include the automated generation of syntactic \cite{AAAI1714603} and semantic \cite{feedback}, \cite{Kaleeswaran:2016:SVF:2950290.2950363}, \cite{Rivers2017} repairs or hints, the automated generation of test cases \cite{datadriventcgen} for judging program correctness, etc.

In this work we focus on test case based feedback in an online education setting, such as on CodeForces \cite{CodeForces}, CodeChef \cite{CodeChef}, LeetCode \cite{LeetCode}, TopCoder \cite{TopCoder}, etc. These popular platforms are usually geared towards students who are proficient in programming but want to hone their algorithm design skills. Thus, the problem of interest in these scenarios is to check whether an incoming submission has a semantic/logical bug, which is typically accomplished by running the submission against a set of test cases. Most often, these test cases are manually designed and require a significant amount of human effort and expertise as it is difficult to anticipate all the errors which may be made by students. Additionally, the online nature of these environments makes temporal efficiency a necessary concern for any automated solution.


We tackle the problem of efficiently and automatically generating a set of quality test cases. We dub our solution, which is inspired by recent advances in symbolic execution-based \cite{DBLP:journals/corr/BaldoniCDDF16} test case generation (specifically \texttt{klee} \cite{klee:paper}) and active learning \cite{activelearning} (for classification), \textbf{A}utomated \textbf{T}esting using \textbf{A}ctive learning and \textbf{S}ymbolic execution (ATAS).

ATAS uses symbolic execution to check the semantic equivalence of a submission with a reference implementation and generates a failing test case if the submission has a logical bug. This method is more accurate than using a hand-designed test suite since it can handle all possible distributions of logical bugs that may be present in the submission. However, as this process can be computationally intensive, ATAS makes novel use of active learning to dramatically reduce the number of submissions for which equivalence checking is performed. In our experiments, ATAS achieves an average speedup of over 2.5x (in terms of runtime) over a baseline (that exclusively performs the aforementioned expensive analysis) without a significant degradation in evaluation accuracy. ATAS reduced the number of expensive program analysis calls by over an order of magnitude, thus yielding a near-optimal speedup in practice over the baseline.
Additionally, ATAS has a parameter which trades off runtime speedup with evaluation accuracy. It can be initialized by the instructors according to their requirements.  

The main contributions of this work are as follows:

\begin{enumerate}
	\item We propose two algorithms to solve the problem of automated evaluation in online setting. The first, Algorithm \ref{baseline}, is based on symbolic execution and is guaranteed to find all buggy (i.e.\ \textit{incorrect}) program submissions (modulo the tool's capability). ATAS (Algorithm \ref{AEM_alg}) is an active learning based augmentation of Algorithm \ref{baseline}.
	\item We analyze the algorithms on many real-world datasets. We found that Algorithm \ref{baseline} is as good in practice as the manually designed high quality test cases. Algorithm \ref{AEM_alg} shows a speedup of $2.5$x over Algorithm \ref{baseline} without a significant degradation in evaluation accuracy.
\end{enumerate}

\section{Related Work}

In our method, we use symbolic execution to check for semantic equivalence. In this process, if a submission semantically differs from the reference implementation, a counterexample is also generated. This counterexample is a test case on which the submission fails i.e.\ its output is different from the reference implementation's output. Thus, our method can be viewed as related to automated generation of test cases such that they catch all the logical bugs for student submissions corresponding to some task. Hence, we briefly discuss the automated test case generation techniques from the literature.

Automated test case generation is an active research topic owing to its utility not just in an online judge setting, but also in testing industrial code. Owing to its practical impact, test case generation for industrial code has been explored in greater detail. We would like to note that there are a few major differences between the scenario of our interest and that of test case generation for industrial code. First, in the industrial code scenario, an assumption that is often made is that most of the source code is correct. This is in contrast to submissions made by students, where errors are much more prevalent. Second, we can assume the availability of a \textit{reference implementation}, a correct implementation of a given algorithmic task, which we leverage in our technique. Such a reference implementation is usually not available in the industrial code setting. Third, computational efficiency is a major concern in the online judge setting, where response times are critical. 

A recent survey \cite{surveytcgen} categorizes the existing test case generation techniques into three types: random-based methods, search-based methods and data mining-based methods. Random-based methods randomly generate a large number of test cases within some defined test case constraints. However, they completely disregard the distribution of bugs that may occur and hence may not generalize well. Search-based methods regard test case generation as an optimization problem and use advanced algorithms such as scatter search, simulated annealing, etc.\ to find the best test set. However, these methods require careful tuning of the fitness function, which manifests itself as an additional computational overhead. Data mining methods usually require a large number of samples -- a requirement which is typically not met in the online judge setting.

In \cite{datadriventcgen}, a recent work, the authors focus on programming assignments for students. They first generate a large number of test cases and then successively refine this set with the help of a base set of submissions, ensuring that the reduced set of tests is capable of discovering all the bugs originally found in the base set. This reduced set of high quality test cases are then used to judge all future submissions. This approach is, however, not directly applicable in the online judge scenario as it is difficult to estimate the ideal number of initial submissions required to form a quality base set of submissions. As their technique does not update itself to accomodate future submissions, it is especially important to set this number correctly. Our experiments show that setting the first 10\% of all submissions aside as the base set is not enough to catch all \textit{incorrect} programs present in the remaining 90\% submissions. 


\section{Background}\label{bg}

Our method ATAS builds upon active learning, symbolic execution and scalable variants of gradient boosting machines. We now briefly describe each of these techniques. 

\textbf{Active learning} \cite{activelearning} is typically used for tasks where we have access to (or a way to generate) a large amount of unlabeled data and a labeling oracle which can be queried to get label of a data sample. Such scenarios may require learning an `accurate' classifier by making as few queries to the oracle as possible. Active learning is an augmentation to usual classification methods which achieves these goals. Uncertainty sampling is an active learning method in which a classifier is successively trained in multiple steps. In each step, a few data samples which are ``confusing'' to the trained classifier are labeled, and then added to the training dataset. This training dataset is then used to retrain the classifier for use in the next step. These steps are repeated till some stopping criterion is satisfied (refer to \cite{activelearning} for more details).

\textbf{Symbolic execution} of the program, unlike concrete execution (which involves executing a program on a single test case), involves running the program with symbolic input variables. This execution assigns a first order boolean formula, $b_s$, to each statement $s$, in the program in such a way that concrete assignments to symbolic inputs which satisfy $b_s$ will also result in the execution of $s$. Symbolic execution has many applications -- one of which is generating a set of test cases that results in the execution of all statements in the program (provided they can be executed by some valid test case). Interested readers can refer \cite{DBLP:journals/corr/BaldoniCDDF16} for more details.

\textbf{Boosting} \cite{Schapire1990} is an ensemble-based method in which many weak classifiers (e.g.\ ``shallow'' decision trees) are successively trained and combined to build a strong classifier. As shown in \cite{GBM}, boosting methods can also be viewed as a ``constrained'' gradient descent-based optimization of expected error for specific loss functions. In \cite{GBM}, the authors build upon this observation and generalize the setting to work for arbitrary differentiable loss functions. The resulting methods are called gradient boosting machines. XGBoost \cite{xgboost:paper} is a scalable variant of these gradient boosted machines.

\section{Labeling Submissions Using Symbolic Execution}\label{pa_klee}

In this section, we use an example to illustrate how symbolic execution can be used to check if a submission is semantically equivalent to reference implementation. For this, we make use of \texttt{klee}, which employs symbolic execution to generate a set of test cases that result in the execution of all reachable statements in program. A statement \textit{s}, is said to be \textit{reachable} if there exists an input to a program which will result in the execution of \textit{s}. Such a set of test cases is said to achieve ``high coverage'' since, put together, they execute all reachable statements of the program.

\begin{figure}[h]
	\centering
	
	\begin{minipage}{0.5\textwidth}
		\raggedleft
\begin{lstlisting}[language=C, caption={Erroneous Submission}, label={lst:incor}, captionpos=b]
#include<stdio.h>

int main() {
  long long int x, ans;
  scanf("%lld", &x);
  ans = x*x*x;
  printf("%lld", ans);

  return 0;
}
\end{lstlisting}
	\end{minipage}
	\begin{minipage}{0.45\textwidth}
		\raggedright
\begin{lstlisting}[language=C, caption={Reference Solution}, label={lst:refsol}, captionpos=b]
#include<stdio.h>

int main() {
  int inp;

  scanf("%d", &inp);
  printf("%d", inp*inp);

  return 0;
}
		\end{lstlisting}
	\end{minipage}
	
\end{figure}

Consider the toy task of taking an integer $i$ as input and computing its square, $i^2$. Furthermore, as is common in most algorithmic tasks, assume that the input integer is constrained to be in the range $1 \le i \le 1000$. Listing \ref{lst:incor} shows an erroneous program submission and Listing \ref{lst:refsol} shows the reference solution. Listing \ref{lst:foranalysis} shows a source file that combines both the erroneous program submission and the reference solution. This combined program has the property that line number $22$ is executed if and only if the reference solution and the erroneous submission vary for some test case. Since \texttt{klee} generates high coverage test cases \cite{klee:paper}, it will try to find a test case that triggers the execution of line number $22$. This results in the generation of a \textit{failing} test case for the erroneous program. We implemented a simple abstract syntax tree (AST) rewriting phase to generate a \textit{combined} program for every submission using the \texttt{pycparser} tool \cite{pycp} in a manner similar to the above illustration.

\begin{lstlisting}[language=C, caption={Code to be analyzed using klee}, label={lst:foranalysis}, captionpos=b]
long long int f1(int inp)
{
  return inp * inp;
  return 0;
}
long long int f2(long long int x)
{
  long long int ans;
  ans = (x * x) * x;
  return ans;
  return 0;
}
int main(){
  int n;
  klee_make_symbolic(&n, sizeof(n), "n");
  klee_assume(1<=n);
  klee_assume(n<=2000);
  long long int a1 = f1(n);
  long long int a2 = f2(n);
  if(a1 != a2)
    klee_assert(0); // Failure if incorrect submission
  return 0;
}
\end{lstlisting}

\section{Automated Evaluation of Student Programs}

In this section, we first formally define the problem of evaluating student programs in an online setting, and then, we explain two algorithms designed to solve it. Each problem instance is a two-tuple $(Q, R)$, where $Q$ is a queue of program submissions for a particular algorithmic task and R is a reference solution to the same task. The output is a two-partition of $Q$ into \textit{A}, the \textit{correct} submissions (i.e.\ those that solve the algorithmic task correctly) and \textit{W}, the \textit{incorrect} submissions (i.e.\ which have some logical error).

Algorithm \ref{baseline} employs the technique described in Section \ref{pa_klee} to label all the submissions correctly -- modulo \texttt{klee}'s capability. 

\begin{algorithm}
	\DontPrintSemicolon
	\KwData{$Q$, a queue of program submissions\\
		$R$, the reference solution }
	\KwResult{$A$ and $W$, the sets of \textit{correct} and \textit{incorrect} programs updated with every new submission}
	$A \longleftarrow \{\}, \:W \longleftarrow \{\}$\;
	$T \longleftarrow \{\} $\tcc{set of klee generated \textit{failing} test cases}\;
		
		\While{$Q$ is not empty}{
			$prog \longleftarrow Q$.pop()\\\;
			\If{ T $\neq \{\} $ and $prog$ fails on some test case t $\in$ T }{
				\tcc{Since it fails, it is definitely incorrect}
				W.add(prog)\;
			}\Else{
			\tcc{label prog using klee and update the sets A, W, T}
			T, A, W $\longleftarrow$ label\_using\_klee($prog, \:R$, T, A, W)
		}
	}
\caption{Baseline\label{baseline}}	
\end{algorithm}

Since the baseline only analyzes those submissions which pass all tests $t \in T$ (Step 10), the set of generated test cases, it will avoid analysis for those submissions which commit a mistake that has been already encountered before. We hypothesize that such redundancy will also be present in the \textit{correct} class i.e.\ there would be many implementations which implement the same solution strategy. ATAS exploits this redundancy to achieve significant speedup over the baseline.

\section{The Proposed ATAS Method }\label{ATAS_motivation}

We use a classifier to characterize the already encountered samples from the \textit{correct} class (details in Section \ref{ATAS_alg}). We then analyze only those submissions for which the classifier is not very confident about its \textit{correct} label. This can be viewed as a variant of uncertainty sampling based active learning approach suited for the online programming setting.

\subsection{ATAS Algorithm}\label{ATAS_alg}

In this section, we describe the ATAS algorithm (Algorithm \ref{AEM_alg}). Since the classifier requires labeled data samples to train on, ATAS works in two phases. First, in the seeding phase, ATAS labels the first \textit{i} submissions using \texttt{klee} and trains a classifier on it (steps 3-7). In the second phase, ATAS processes the submissions in an online fashion.

The second phase of ATAS largely resembles the baseline, but has one important difference. ATAS checks for and eliminates the expensive analysis of suspected \textit{correct} samples i.e.\ those that classifier confidently labels as \textit{correct} (Steps 14-19). This speeds up ATAS since the redundant analysis of already encountered \textit{correct} implementation strategies is avoided. Our experiments back this up by demonstrating that ATAS performs far fewer \texttt{klee} calls without significantly degrading evaluation accuracy on many real-world datasets. To update the classifier with newly encountered implementation strategies used by the incoming \textit{correct} samples, we retrain the classifier with the updated labeled samples at regular intervals (Steps 21-22).


We now discuss our method for selecting the parameter $thresh$. The value of $thresh$ defines which samples are ``confusing'' to the classifier and require \texttt{klee} for labeling (Step 15). For this, while training the classifier, we set aside $20\%$ of the labeled training data as a validation set. We then train the classifier on the remaining $80\%$, and evaluate the trained classifier on it. We set the value of $thresh$ to the least threshold resulting in a false positive rate below $F$.

\subsection{Program Representation}

We view the program as a sequence of tokens (specifically, tokens identified by \texttt{pycparser}) and choose n-grams as the feature vocabulary. We first anonymize all identifiers present in the program. This is done because different submissions sharing the same solution strategy may use different identifiers (function and variable names, etc.). We extract all the n-grams present in first \textit{i} programs and use them as our feature set. A program is then encoded as a bag of all the n-grams present in its text. In our experiments, we found that $n = 3$ was a good choice for all the datasets. For example, suppose that the ordered set of 3-gram features is $\{abc, bcd, cde\}$ and suppose that the program token sequence is $abcd$. Then the program is represented as the bag: $\{abc, bcd\}$ and is encoded as the vector $\begin{matrix}\begin{pmatrix} 1 & 1 & 0 \end{pmatrix}\end{matrix}$.

\begin{algorithm}
	\DontPrintSemicolon
	\KwData{$Q$, a queue of program submissions\\
		$i$, number of initial programs to be labeled using \texttt{klee}\\
		$r$, number of programs after which classifier is retrained\\
		$C$, an instance of chosen classifier family \\
		$F$, maximum allowable false positive rate on the validation split of training data \\
		$R$, the reference solution }
	\KwResult{$A$ and $W$, set of \textit{correct} and \textit{incorrect} submissions updated with every new submission}
	A\textsubscript{klee} $\longleftarrow \{\}$ \tcc{\texttt{klee} labeled $correct$}
	$A \longleftarrow \{\}, \:W \longleftarrow \{\} $\;
	$T \longleftarrow \{\} $\tcc{set of \texttt{klee} generated \textit{failing} test cases}\;
	\tcc{label first $i$ submissions using \texttt{klee} and train $C$ on the labeled data}
	\For{j $= 1, 2, \dots i$ }{
	T, A\textsubscript{klee}, W $\longleftarrow$ label\_using\_klee(Q.pop(), R, T, A\textsubscript{klee}, W)
	}
	A $\longleftarrow$ A $\cup$ A\textsubscript{klee}\;
	features $\longleftarrow$ generate\_features(A$\cup$ W) \\
	C, thresh $\longleftarrow$ train\_and\_get\_thresh(C, $F$, A\textsubscript{klee}, W, features) \\
	\tcc{Process the remaining programs online}
	\While{$Q$ is not empty}{
			$prog \longleftarrow Q$.pop()\\\;
			\If{T $\neq \{\} $ and $prog$ fails on some test case t $\in$ T }{
				W.add($prog$)\;
				}\Else{
				\tcc{probability of it being \textit{correct} as per the $C$}
				repr $\longleftarrow$ encode\_program(prog, features)\;
				$prob \longleftarrow C.$predict\_probability(repr)\\
				\If{$prob < thresh$}{
					T, A\textsubscript{klee}, W $\longleftarrow$ label\_using\_klee($prog$, R, T, A\textsubscript{klee}, W)\;
					A $\longleftarrow$ A $\cup$ A\textsubscript{klee}
				}\Else{A.add(prog)}
				}\;
				
				\If{$\left| A \cup W \right| - i$ is a multiple of $r$}{
				C, thresh $\longleftarrow$ train\_and\_get\_thresh(C, $F$, A\textsubscript{klee}, W, features)
					}
		}

\caption{Online ATAS algorithm\label{AEM_alg}}
\end{algorithm}


\subsection{Classifier Family}

Since the features are categorical, we made use of decision trees and gradient-boosted decision trees (specifically XGBoost \cite{xgboost:paper} because of its scalability) as our classifiers. These classifiers are known to be well-suited for data having categorical features. We also use \texttt{k-nn} classifiers in our experiments since they are easy to train without requiring much hyperparameter tuning.

\section{Experiments}\label{experiments}

In this section, we summarize the results of our experiments. The experiments are designed to analyze: 1) How precise is the baseline when compared with CodeForces' categorization?, 2) How fast is ATAS over the baseline?, 3) How precise is ATAS when compared with the baseline?.

To analyze the speedup achieved by ATAS over the baseline, we choose two parameters, the number of submissions which are analyzed using \texttt{klee} (henceforth \texttt{klee} \textit{calls}) and the runtime. The number of \texttt{klee} calls made is an important metric, as the \texttt{klee}-based analysis can be computationally intensive and hence time consuming. All our experiments are performed on an Intel(R) Xeon(R) E5-1620 4-core 8-thread machine with 24GB of RAM. The algorithms are implemented to make use of all 8 threads. Also, we choose a \texttt{klee} timeout of $15$ seconds. If no failing test cases are generated at the end of \texttt{klee}'s analysis, we assume the submission to be correct.

\subsection{Dataset Collection and Generation of Combined Programs}

\begin{wrapfigure}{l}{0.6\textwidth}
	\begin{tikzpicture}[thick,scale=0.6]
	\begin{axis}[
	scale only axis,
	legend columns=2,
	legend style={
		/tikz/column 2/.style={
			column sep=10pt,
		},
	},
	xmin=1, xmax=10,
	ymin=0, ymax=110,
	ylabel={\textit{correct} submissions(\%)},
	xlabel={Deciles (sorted in time)},
	label style={scale=1.7},
	ytick={20, 40, 60, 80, 100},
	xtick={1, 2, 3, 4, 5, 6, 7, 8, 9, 10}
	]
	\addplot[mark=*, color=blue] 
	table[x=Y,y=X] {dist_data/bshovel.dat};
	\addplot[mark=o, color=black] 
	table[x=Y,y=X] {dist_data/buttons.dat};
	\addplot[mark=square*, color=red, dotted] 
	table[x=Y,y=X] {dist_data/insomnia_cure.dat};
	\addplot[mark=triangle*, color=green] 
	table[x=Y,y=X] {dist_data/game_sticks.dat};
	\addplot[mark=otimes*, color=cyan, dashed] 
	table[x=Y,y=X] {dist_data/solbanana.dat};
	\addplot[mark=diamond*, color=purple] 
	table[x=Y,y=X] {dist_data/watermelon.dat};
	\legend{D1, D2, D3, D4, D5, D6}
	\end{axis}
	\end{tikzpicture}
	\caption{Figure showing percentage of \textit{correct} submissions (sorted in time) in various deciles.}
	\label{figure_dist}
\end{wrapfigure}
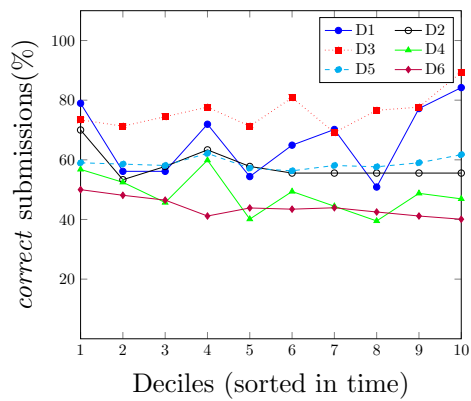

We chose a few problems from CodeForces and downloaded all the corresponding \texttt{C} submissions. We restricted ourselves to problems which require a few well-defined integer inputs (less than 5) and output only a single string or integer. It may seem a bit restrictive but we note that first, there are a large number of problems on CodeForces which satisfy these constraints. Second, one can write a more complicated AST rewrite phase to handle problem statements with more general inputs and hence, ATAS is in no way restricted by the aforementioned constraints. The submissions in our dataset consists of loops, conditionals, \texttt{switch} statements etc.\ and hence represent a rich class of program submissions.

For each chosen problem statement, we downloaded all the submissions' code, time stamp, username and the CodeForces' verdict, i.e.\ one of \texttt{accepted} (\textit{correct}) or \texttt{wrong answer} (i.e.\ \textit{incorrect}). We did not download the submissions having other verdicts such as Time Limit Exceeded, Memory Limit Exceeded, Compilation Error, etc. We then removed all the submissions defining functions other than \texttt{main}, declaring arrays with size $> 1000$, making use of external function invocations, containing \texttt{printf} statements with invalid format specifiers, etc. Table \ref{table_datasets} shows the details of the collected dataset and Fig. \ref{figure_dist} shows the distribution of classes in the time-sorted data samples. From Fig. \ref{figure_dist} we can note that all the \textit{incorrect} submissions are not skewed in the initial phase.

\begin{table}[H]
	\centering
	\caption{Datasets}
	\begin{tabular}{| P{0.41\textwidth} | P{0.15\textwidth} P{0.15\textwidth} P{0.12\textwidth} P{0.12\textwidth}|}
		\hline
		Dataset Name & \multicolumn{4}{ P{0.54\textwidth} |}{Pruned Dataset's details}\\
		\hline
		& Total & \textit{correct} & \textit{incorrect} & \# features\\
		\hline
		Buy A Shovel (D1) & 647 & 426 & 221 & 444 \\
		Buttons (D2) & 1000 & 579 & 421 & 579\\
		Insomnia Cure (D3) & 1056 & 800 & 256 & 522 \\
		Game With Sticks (D4) & 1815 & 877 & 938 & 620\\
		Soldiers and Banana (D5) & 2472 & 1453 & 1019 & 760\\
		Watermelon (D6) & 16470 & 7259 & 9211 & 1538 \\
		\hline
	\end{tabular}
	\label{table_datasets}
\end{table}

	
	

\subsection{Setup}\label{setup}

ATAS, in addition to the choosen classifiers' hyperparameters, also has its own hyperparameters -- namely $i$, the number of initial submissions to be labeled using \texttt{klee}; $r$, the number of submissions after which the classifier is to be retrained and $F$, the maximum allowable false positive rate. Choosing $i$ and $r$ is specific to the problem statement for which ATAS is deployed. After experimenting, we found that 10\% of the total data is a good value for both. We would like to note that when ATAS is deployed, the instructor may not have access to the full data (i.e.\ all the submissions made by the students). However, the instructors may set an approximate value based on statistics of submissions belonging to tasks of a similar difficulty. 

For \texttt{k-nn} classifiers we found a value of $k = 6$ to work well in practice. For decision tree classifiers, we do a random search to obtain ten hyperparameter configurations during training and choose the one with the best validation accuracy. We use the \texttt{Scikit-learn} \cite{sklearn} package to implement the two classifiers. For \texttt{XGBoost}, we choose $max\_depth = 7$ and $n\_estimators=100$.

To check our method's generalization ability, between each retraining phase, we keep out $10\%$ of the data as test data and use remaining $90\%$ data (henceforth, \textit{comparison data}) for comparison with baseline. Whenever we retrain the classifier, we calculate the trained classifier's precision and recall on the next batch's test data. The results are shown in Table \ref{table_precrecall}. In the rest of this section, all the results are mentioned for the \textit{comparison data}.

\begin{table}[H]
	\centering
	\caption{ATAS + XGBoost's precision and recall on different datasets}
	\begin{tabular}{| P{0.45\textwidth} | P{0.2\textwidth} P{0.2\textwidth } |}
		\hline
		Datasets & Avg. Precision & Avg. Recall\\
		\hline
		D1 & 0.83 & 0.78\\
		
		D2 & 0.96 & 0.82 \\
		
		D3 & 0.87 & 0.85\\
		
		D4 & 0.84 & 0.85 \\
		
		D5 & 0.88 & 0.81 \\
		
		D6 & 0.78 & 0.98 \\
		\hline
	\end{tabular}
	\label{table_precrecall}
\end{table}

\subsection{Results} 

First, we check how precise baseline is as compared to CodeForces. For this, we compare the inaccuracies in baseline labeling with CodeForces' labeling. Since \textit{incorrect} submission implies that there is a test case for which the submission fails, \textit{error} refers to the submissions which are actually \textit{incorrect} but are marked as \textit{correct} by the corresponding algorithm compared. The aggregate results of all datasets are shown in Table \ref{table_labelnoise}. Clearly, the baseline can be seen to perform as good as CodeForces. 


We next wish to check how fast ATAS is in comparison to the baseline. For this, we need to choose the value of $F$ and also our preferred classifier family. We compared our chosen classifier families for different values of $F$. We concluded that XGBoost is the preferred classifier due to its superior speedup across all datasets, with no significant difference in evaluation accuracy. We summarize our findings by first showing the behavior of ATAS with XGBoost for various values of $F$ in Table \ref{table_xgbconf}. The results demonstrate that $F$ controls the trade-off between speedup and evaluation accuracy, with higher values of $F$ favoring speedup over error. We recommend using a value of $F=0.3$. Figure \ref{figure_ATASxgbBLN} shows the runtime comparison of baseline and ATAS with XGBoost ($F=0.3$) for various datasets.

\begin{minipage}{0.45\textwidth}
	\raggedleft
	\begin{table}[H]
		\caption{Error by different labeling oracles}
		\begin{tabular}{| P{0.6\textwidth} | P{0.3\textwidth}|}
			\hline
			Labeling Oracle & Error\\
			\hline
			CodeForces & 124 \\
			baseline & 104\\
			\hline
		\end{tabular}
		\label{table_labelnoise}
	\end{table}
\end{minipage}
\begin{minipage}{0.45\textwidth}
	\raggedright
	\begin{figure}[H]
		\centering
		\begin{tikzpicture}[thick,scale=0.45]
		\begin{axis}[
		ybar stacked,
		enlargelimits=0.15,
		legend style={at={(0.5,-0.10)},
			anchor=north,legend columns=-1},
		ylabel={Time(minutes)},
		label style={scale=1.7},
		symbolic x coords={D1, D2, D3, D4, D5, D6},
		xtick=data,
		x tick label style={rotate=45,anchor=east},
		]
		\addplot+[ybar] plot coordinates {(D1,6.55) (D2, 7.51) 
			(D3, 9.09) (D4, 10.84) (D5, 23.16) (D6, 25.39)};
		\addplot+[ybar, postaction={pattern = north east lines}] plot coordinates {(D1, 5.78) (D2, 12.13) 
			(D3, 20.32) (D4, 18.63) (D5, 52.32) (D6, 35.15)};
		\legend{ATAS with XGBoost, Baseline}
		\end{axis}
		\end{tikzpicture}
		\caption{Cumulative time comparison of ATAS with XGBoost against Baseline}
		\label{figure_ATASxgbBLN}
	\end{figure}
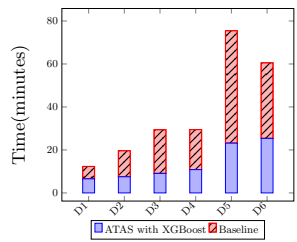
\end{minipage}

\begin{table}
	\caption{Comparison of XGBoost with different values of $F$. Speedup is the ratio of baseline's runtime and corresponding algorithm's runtime. Error is the number \textit{incorrect} samples labeled as \textit{correct} out of the total \textit{incorrect} samples as per the baseline. We note that higher value of $F$ favors speedup at the expense of error.}
	\begin{tabular}{| P{0.21\textwidth} | P{0.12\textwidth} P{0.12\textwidth} | P{0.12\textwidth} P{0.12\textwidth} | P{0.12\textwidth} P{0.12\textwidth} |}
		\hline
		Datasets & \multicolumn{2}{ P{0.2\textwidth} |}{$F=0.1$} & \multicolumn{2}{ P{0.2\textwidth} |}{$F=0.3$} & \multicolumn{2}{ P{0.2\textwidth} |}{$F=0.5$}\\
		\hline
		
		& Speedup & Error & Speedup & Error & Speedup & Error \\
		\hline
		
		D1 & 1.26 & \textbf{1}/182 & 1.88 & 2/182 & \textbf{2.45} & 5/182\\
		
		D2 & 1.68 & \textbf{0}/375 & 2.61 & 1/375 & \textbf{3.04} & 1/375 \\
		
		D3 & 1.68 & \textbf{3}/235 & \textbf{3.23} & 5/235 & 3.05 & 6/235\\
		
		D4 & 1.54 & \textbf{0}/834 & 2.71 & 1/834 & \textbf{3.25} & 3/834\\
		
		D5 & 1.53 & \textbf{2}/908 & 3.25 & 2/908 & \textbf{3.91} & 3/908\\
		
		D6 & 1.77 & \textbf{6}/8333 & 2.38 & 13/8333 & \textbf{2.72} & 13/8333 \\
		\hline
	\end{tabular}
	\label{table_xgbconf}
\end{table}

We now compare all the chosen classifier families with $F$ set to $0.3$. The results are shown in Table \ref{table_XKnnDTCVcomparison}. We observe that overall, ATAS with XGBoost is faster than ATAS with other classifier families, without a significant loss in evaluation accuracy. Table \ref{table_XKnnDTCVkleeCalls} shows the comparison of these configurations to the baseline in terms of the number of \texttt{klee} calls. Each configuration has two columns. The first column, \textit{Total}, shows the total number of \texttt{klee} calls made by the configuration while the second column (\textit{Last 90\%}) shows the number of \texttt{klee} calls made by the configuration compared to those made by the baseline for the last $90\%$ of the data (i.e.\ when the classifier comes into the picture). We note that ATAS with XGBoost clearly outperforms all the other configurations by a huge margin. We see that for the remaining $90\%$ of the data, ATAS with XGBoost makes 1-2 orders of magnitude smaller number of \texttt{klee} calls than the baseline, thus clearly demonstrating the usefulness of our technique.

\vspace{-0.5cm}
\begin{table}
	\caption{Comparison of XGBoost, \texttt{k-nn} classifier and decision tree for $F=0.3$ (notations are same as used in Table \ref{table_xgbconf}).}
	\begin{tabular}{| P{0.21\textwidth} | P{0.12\textwidth} P{0.12\textwidth} | P{0.12\textwidth} P{0.12\textwidth} | P{0.12\textwidth} P{0.12\textwidth} |}
		\hline
		Dataset & \multicolumn{2}{ P{0.24\textwidth} |}{\texttt{k-nn}} & \multicolumn{2}{ P{0.24\textwidth} |}{XGBoost} & \multicolumn{2}{ P{0.24\textwidth} |}{Decision Tree}\\
		\hline
		& Speedup & Error & Speedup & Error & Speedup & Error\\
		\hline
		D1 & 1.42 & 4/182 & \textbf{1.88} & 2/182 & 1.73 & \textbf{1}/182\\
		
		D2 & 1.81 & \textbf{0}/375 & \textbf{2.61} & 1/375 & 1.88 & \textbf{0}/375 \\
		
		D3 & 1.45 & \textbf{4}/235 & \textbf{3.23} & 5/235 & 2.59 & 5/235\\
		
		D4 & 1.69 & 3/834 & \textbf{2.71} & \textbf{1}/834 & 1.76 & 3/834\\
		
		D5 & 1.87 & \textbf{2}/908 & \textbf{3.25} & \textbf{2}/908 & 2.13 & \textbf{2}/908\\
		
		D6 & 0.81 & 5/8333 & \textbf{2.38} & 13/8333 & 0.97 & \textbf{2}/8333\\
		\hline
	\end{tabular}
	\label{table_XKnnDTCVcomparison}
\end{table}

\begin{table}[H]
\caption{Comparison of XGBoost, \texttt{k-nn} and decision tree classifier for the number of \texttt{klee} calls made.} 
	\begin{tabular}{| P{0.14\textwidth} | P{0.12\textwidth} | P{0.1\textwidth} P{0.12\textwidth} | P{0.1\textwidth} P{0.12\textwidth} | P{0.1\textwidth} P{0.12\textwidth} |}
		\hline
		Dataset & baseline & \multicolumn{2}{ P{0.22\textwidth} |}{\texttt{k-nn}} & \multicolumn{2}{ P{0.22\textwidth} |}{XGBoost} & \multicolumn{2}{ P{0.22\textwidth} |}{Decision Tree}\\
		\hline
		& Total & Total & Last 90\% & Total & Last 90\% & Total & Last 90\% \\
		\hline
		D1 & 407 & 158 & 101/351 & \textbf{150} & \textbf{93}/351 & 154 & 97/351\\
		
		D2 & 530 & 158 & 68/458 & \textbf{145} & \textbf{55}/458 & 216 & 126/458 \\
		
		D3 & 719 & 327 & 233/639 & \textbf{140} & \textbf{46}/639 & 192 & 98/639\\
		
		D4 & 804 & 290 & 128/692 & \textbf{205} & \textbf{43}/692 & 354 & 192/692\\
		
		D5 & 1319 & 496 & 274/1175 & \textbf{313} & \textbf{91}/1175 & 514 & 292/1175\\
		
		D6 & 6496 & 2170 & 688/5752 & \textbf{1530} & \textbf{48}/5752 & 2265 & 783/5752 \\
		\hline
	\end{tabular}
	\label{table_XKnnDTCVkleeCalls}
\end{table}

\subsection{Discussion and Future Work}

In the previous section we have compared the baseline and CodeForces on the parameter of \textit{error} (i.e. the inaccuracies in labeling). The symbolic execution based baseline has clearly showed a promise since it has found few \textit{incorrect} (124 in total) submissions which are marked \textit{correct} by the hand designed test case based labeling of CodeForces. However, the baseline also missed few \textit{incorrect} submissions (104 in total). Overall, we can say that the baseline is competitive with CodeForces on this parameter.

ATAS is compared to the baseline on the parameters of \textit{error} and temporal efficiency (\textit{speedup} and \texttt{klee} calls). Since the CodeForces test cases are hand designed, it cannot be evaluated with the proposed algorithms in terms of temporal efficiency. The results show that ATAS with XGBoost, for recommended hyperparameters, has a small degradation in terms of \textit{error}. Specifically, it made 24 inaccurate labelings on a real world dataset consisting a total of about 21K submissions (comparison data i.e.\ 90\% of total data). Since the baseline and CodeForces have similar performance, we can see that ATAS with XGBoost performs quite competitively against both of them. In terms of both the parameters of temporal efficiency, ATAS with XGBoost has shown a dramatic improvement over the baseline.

In the present form, both of our methods are implemented for programs which take a few inputs (less than 5) and produce only a single string or integer as output. It also has few other restrictions mentioned in Section \ref{setup}. As a part of future work, we intend to extend AST rewrite phase and evaluate our system for other more challenging programming tasks. Also, we intend to perform a user study to evaluate the ATAS for pedagogical limitations that might be introduced because of slight degradation in accuracy.

\section{Conclusion}

In this work we proposed a solution to the problem of automatically evaluating student programs in an online setting. Our solution uses symbolic execution to evaluate the student programs, thus providing more confidence than the hand-designed test case-based solution in use today. It uses active learning to reduce the number of submissions for which the symbolic execution based analysis is required. We achieve an average speedup of $2.5$x over a baseline that makes use of only symbolic execution based analysis on a real-world dataset comprising of 6 tasks, without significant drop in evaluation accuracy.

%
%
\bibliographystyle{splncs}
\bibliography{references}
	
\end{document}